\documentclass{webofc}
\usepackage{fnpos}

\usepackage[varg]{txfonts}   
\usepackage{hyperref}
\usepackage{url}

\newcommand{\ptor}{\texttt{p2r}\xspace}
\newcommand{\ptorSpack}{\texttt{p2r-spack}\xspace}
\hypersetup{colorlinks=true,citecolor=blue,urlcolor=blue,linkcolor=blue}
\begin{document}
\title{Packaging HEP Heterogeneous Mini-apps for Portable Benchmarking and Facility Evaluation on Modern HPCs}

\author{\firstname{Mohammad} \lastname{Atif}\inst{1}\fnsep\thanks{\email{fmohammad@bnl.gov}} \and
       \firstname{Pengfei} \lastname{Ding}\inst{2}\fnsep\thanks{\email{pding@lbl.gov}} \and
       \firstname{Ka Hei Martin} \lastname{Kwok}\inst{3}\fnsep\thanks{\email{kkwok@fnal.gov}} \and
       \firstname{Charles} \lastname{Leggett}\inst{2}\fnsep\thanks{\email{cgleggett@lbl.gov}}
}

\institute{
Brookhaven National Laboratory, Upton, NY 11973, USA \and 
Lawrence Berkeley National Laboratory, Berkeley, CA 94720, USA \and
Fermi National Accelerator Laboratory, Batavia, IL 60510, USA 
 }

\abstract{
High Energy Physics (HEP) experiments are making increasing use of GPUs and GPU dominated High Performance Computer facilities. Both the software and hardware of these systems are rapidly evolving, creating challenges for experiments to make informed decisions as to where they wish to devote resources. In its first phase, the High Energy Physics Center for Computational Excellence (HEP-CCE) produced portable versions of a number of heterogeneous HEP mini-apps, such as \ptor, FastCaloSim, Patatrack and the WireCell Toolkit, that exercise a broad range of GPU characteristics, enabling cross platform and facility benchmarking and evaluation. However, these mini-apps still require a significant amount of manual intervention to deploy on a new facility.

We present our work in developing turn-key deployments of these mini-apps, where by means of containerization and automated configuration and build techniques such as Spack, we are able to quickly test new hardware, software, environments and entire facilities with minimal user intervention, and then track performance metrics over time.
}
\maketitle
\section{Introduction}
\label{intro}
High energy physics experiments continue to increase their usage of HPC resources,  with diverse workloads across a range of applications such as event generation, simulation, reconstruction and analysis, using both CPU and GPU hardware. However, deploying applications to HPCs is not as trivial as sending a job to generic grid hardware which tend to have much more uniform software and hardware stacks. HPCs are like snowflakes - each is unique at both the hardware and software levels, and so extensive modifications of the applications are usually required. Sometimes the process is so onerous, that it might not be worth the effort.

Thus it is important to be able to quickly evaluate systems for suitability for various workflows, and to also track system performance with known benchmarks to understand the effects of hardware and software changes. While standard HPC benchmarks like High Performance Linpack or MLPerf exist, they do not readily extrapolate to performance of HEP specific applications. Thus it is critical to be able to run application benchmarks that are representative of real HEP workloads. We have extracted a number of such applications from HEP experiments such as CMS, ATLAS and DUNE, and now the challenge is to package them in a way that enables quick deployment by non-experts at a variety of facilities.

Tracking both deployment success and application performance is very useful for monitoring a system as it evolves and changes are made to both software and hardware. Library incompatibilities or API changes often appear over time, and without regular monitoring can catch users by surprise when say an urgent production run is required. Regular, automated builds and benchmarks will help both users and facilities identify problems as they emerge.

There are two obvious choices for packaging technologies - containerization and Spack. The first allows a self-contained deployment, but depending on the software dependencies, may require large base images, with different images required for different hardware backends (eg NVIDIA vs AMD GPUs). The second allows for the distribution of just the source code, but depending on what packages are available on site, may require significant compilation. We will explore using these two technologies with the \ptor application for Spack, and FastCaloSim for containerization. 

\section{Packaging \ptor Using Spack}
In collider experiments, reconstructing the tracks of charged particles propagating through the detectors under a magnetic field is one of the most arithmetically intensive task.
\ptor~\cite{p2r} is a lightweight mini-app that performs the core computations involved in track reconstruction, specifically propagating the track state to the next layer of detector in the radial direction and performing the Kalman update~\cite{Kalman} at the new layer. 

\ptor has been implemented with various portability technologies that covers all three major GPU vendors, as well as multi-threaded CPU reference implementation. 
The kernel of \ptor is abstracted sufficiently that it has minimal dependencies on other software packages or experimental framework. 
The wide coverage of platforms and minimal dependencies make \ptor a prime target for portable benchmarking and evaluation. 

Spack~\cite{spack} is chosen to be the packaging technology for \ptor primarily for its wide adoption in the HPC computing community. Its simplicity and flexibility for versioning and configurations are particularly suitable for integrating into the HPC computing environment. As the GPU portability technologies used in \ptor are rapidly evolving, the flexibility of Spack will allow \ptorSpack to be evaluated easily through these changes. 
\subsection{Spack implementation}
Prior to this work, \ptor did not have a uniform build system for the range of implementations, partly due to the diverse needs of compilers for various backends. The first step for adopting to Spack is to create a build system for \ptor using one that Spack supports, which is chosen to be CMake. 
As a CMake package, each \ptor implementations is organized into individual sub-directories, containing the detailed compilation options specific to the implementation, while the top-level directory contains the controls of run parameters common to all implementations.

Interfacing to \ptorSpack, the run parameters, such as number of events and number of tracks, are fully exposed to users such that users can adjust the problem size suitable for each evaluation. 
The \ptor implementations and backends are controlled with the \texttt{variants} feature of Spack. 
Two multi-valued \texttt{variants} options are used to specify the implementations~(\texttt{cuda}, \texttt{kokkos}, \texttt{alpaka}, \texttt{alpaka}, \texttt{stdpar}, \texttt{sycl}) and backend (\texttt{nvidia}, \texttt{amd}, \texttt{cpu}).
Detailed compiler options in each implementation and backend combination are hidden from for simplicity and reproducibility.

\begin{figure}
    \centering
    \includegraphics[width=0.5\linewidth]{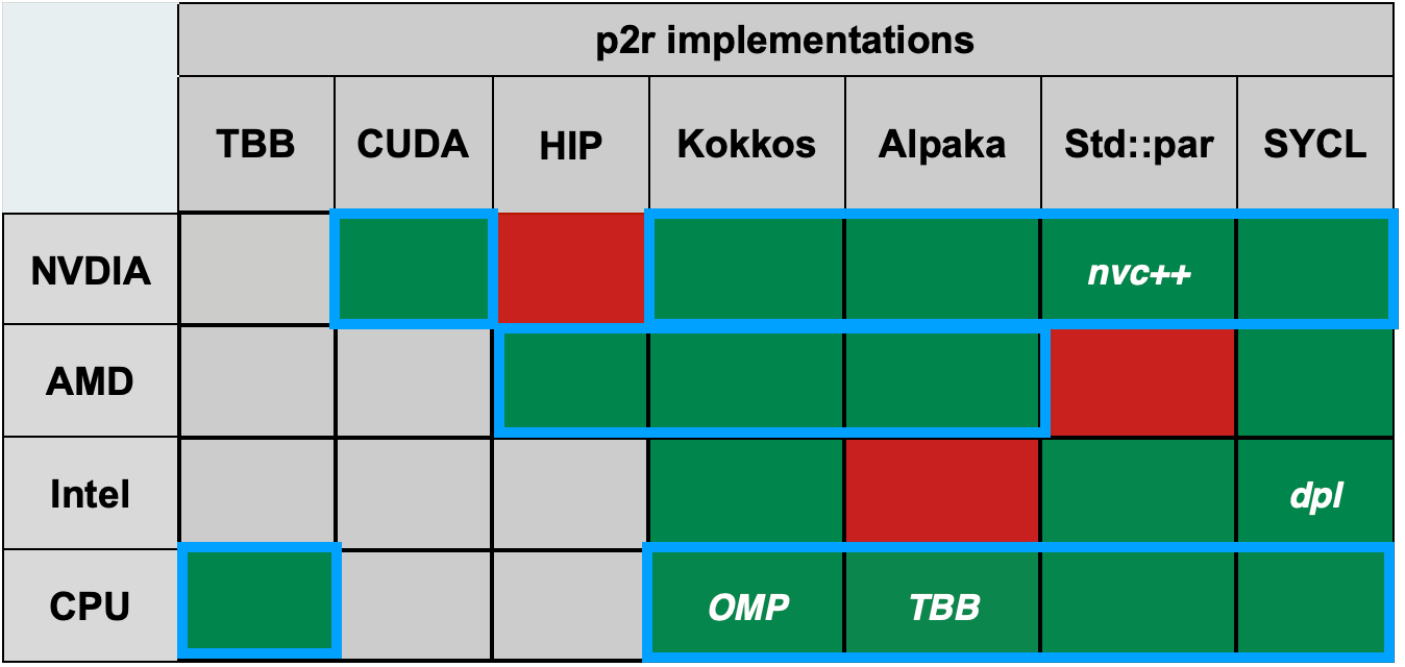}
    \caption{Available \ptor implementations on different execution backends (Green solid cells). This work implements the Spack installation method of the NVIDIA and CPU backends for all \ptor implementations, and AMD backends for HIP, Kokkos and Alpaka implementation~(Blue boarded cells).}
    \label{fig:p2r}
\end{figure}

Figure~\ref{fig:p2r} shows the available \ptor implementations and the currently supported combination of implementations and backends using Spack.

\subsection{External packages}
\ptor depends on vendor software and compilers, such as \texttt{CUDA} and \texttt{nvc++} or portability libraries, such as Kokkos and Alpaka, for specific implementations. 
Obtaining a consistent set of these software with compatible versions is a difficult task, especially across different platforms, and should be factored out of the \ptor package.
To address this issue, a Spack package can specify the compatible range of required software and the users can provide these external packages from system-installed locations.
This approach avoids installing large, duplicated software from source via Spack while satisfying the package requirements.  

Instructions to specify these external packages for \ptor is listed on the \ptor repository~\cite{ptorSpack}.

Overall, packaging with Spack provided a much easier path to deploy \ptor across different HPC platforms. Multi-step manual installation are simplified into a single command. 
Given the wide coverage of \ptor across portability technologies and GPU hardware, we envision \ptorSpack to be a useful tool for benchmarking performance metrics on various modern HPCs for HEP. 

\section{Packaging FastCaloSim Using Containerization and Automating benchmarking with Continuous Integration tools}

FastCaloSim performs a parametrized simulation of the ATLAS Liquid Argon Calorimeter~\cite{FastCaloSim}, which allows for much faster simulation than the standard Geant4 method. Originally written in C++ for CPUs, it has been ported~\cite{FCS_port} to CUDA~\cite{CUDA}, HIP~\cite{HIP}, Kokkos~\cite{Kokkos_1,Kokkos_2}, Alpaka~\cite{Alpaka}, SYCL~\cite{SYCL}, OpenMP~\cite{OpenMP} and std::par~\cite{stdpar}, and can run on NVIDIA, AMD and Intel GPUs as well as multicore CPUs. All of these ports are kept within the same branch of a single git repository. The package uses CMake as a build tool, and the desired backend is specified using flags at configuration time, eg \verb|cmake -DUSE_KOKKOS=ON| or \verb|cmake -DUSE_STDPAR=ON -DSTDPAR_TARGET=GPU|. Depending on which port is selected, dependency libraries such as Kokkos or Alpaka may also need to be installed on the system. The other main library dependency is ROOT, which FastCaloSim uses to read its input data files.

To facilitate the deployment and execution of FastCaloSim across different computing platforms, Docker containers have been developed using base images for NVIDIA and AMD GPUs. Support for various portability libraries as mentioned above has been incorporated as needed, ensuring compatibility with various architectures. Each container includes ROOT v6.32.02, which is compiled with the appropriate toolchain to match the target hardware.

        \begin{figure}
            \centering
            \includegraphics[width=0.95\linewidth]{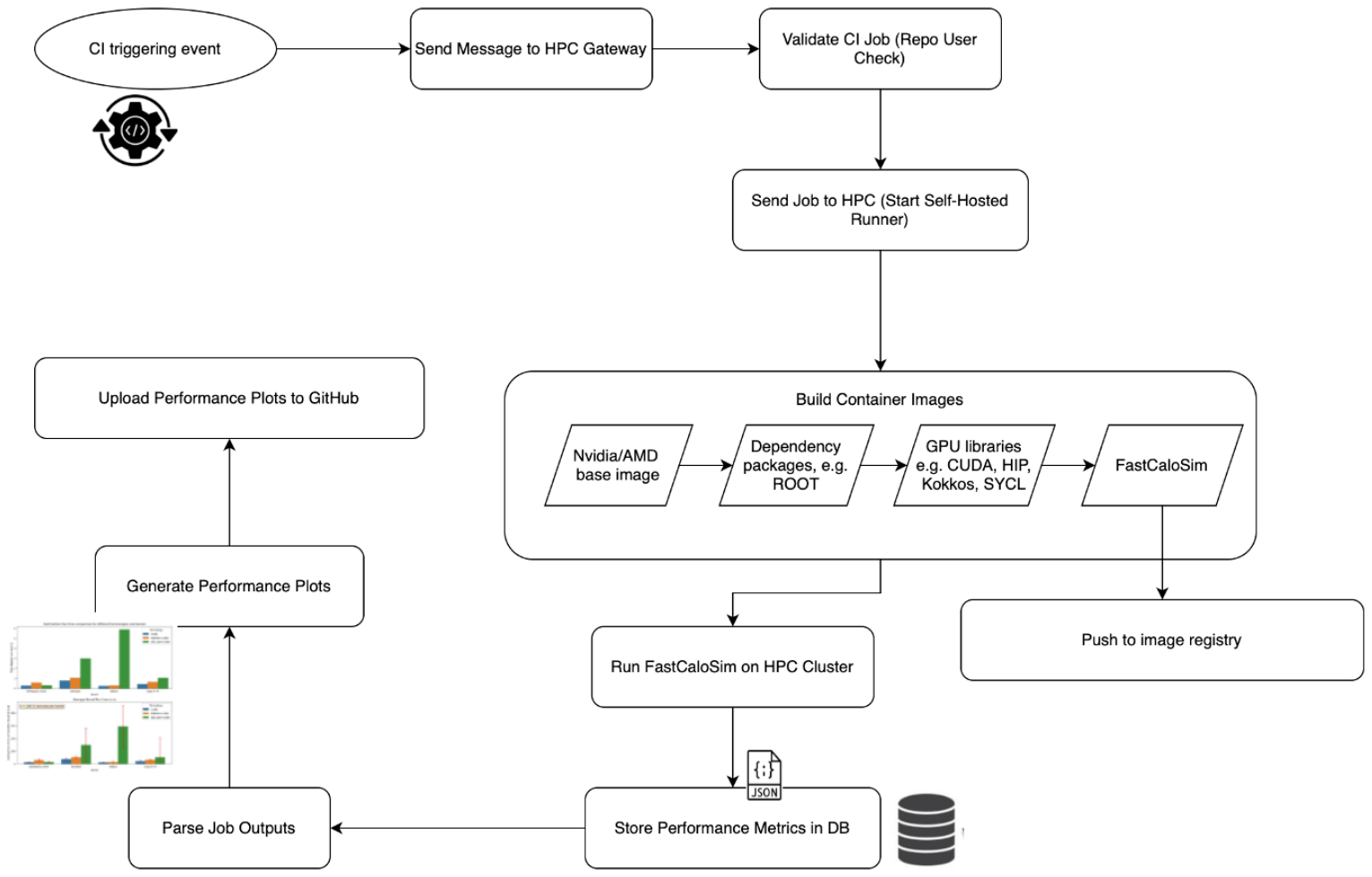}
                \caption{The CI pipeline for containerizing and benchmarking FastCaloSim.}
            \label{fig:fcs-pipeline}
        \end{figure}

Continuous Integration (CI) workflows have been implemented to automate the testing and benchmarking of FastCaloSim as illustrated in Figure~\ref{fig:fcs-pipeline}. These workflows are triggered through GitHub Actions, which initiate a sequence of automated processes whenever a relevant repository update occurs. When a CI workflow is triggered, a GitHub webhook sends a message to an HPC gateway, which serves as an intermediary for job execution on HPC resources. The gateway first verifies whether the job originates from an allowed repository and branch and ensures that the GitHub user has a valid account with the HPC facility. Upon successful validation, the gateway dispatches the job to the HPC system by launching a self-hosted runner remotely. This runner executes the designated CI job and terminates itself immediately upon completion, optimizing resource utilization.

The job execution sequence begins by downloading the appropriate base Docker image, which serves as the foundation for building the FastCaloSim repository with all necessary dependencies. Once the build process is complete, the resulting container image is tagged and pushed to an image registry, making it available for subsequent deployments. The FastCaloSim binary is then executed with various particle and energy configurations to assess performance across different simulation scenarios. The outputs generated during execution are parsed to extract key performance metrics, which are subsequently uploaded to a centralized database. These performance metrics, along with detailed system specifications, are stored for analysis, and automated scripts generate plots to visualize performance trends over time.

This integration of containerized execution and automated validation not only enhances the reliability of FastCaloSim but also streamlines the benchmarking workflow by ensuring consistent performance evaluations across different platforms. By leveraging containerization and CI, benchmarking can be conducted in a reproducible and automated manner, allowing for systematic performance tracking from local development environments to large-scale HPC facilities.

\section{Conclusions}

Both Spack and Docker containerization offer paths for rapid deployment and benchmarking of applications. Containerization has the benefit of allowing the packaging of the full software stack, without needing any external libraries to exist on the target system, but at the price of large container images if many dependencies are required. Spack offers a smaller footprint, and easier customization of used dependencies, allowing different versions to be quickly tested if they are already present on the target system. However adding Spack to an existing project can require considerable extra effort if it is not already compatible with a Spack supported build system, and comes with a greater risk of build time errors if incompatible library versions are present on the system, or of long and complicated builds if a full dependency tree needs to be downloaded and built from scratch. While both Spack and containerization work well for the test cases that were studied, more complicated packages with large dependency trees would more suitable for a containerized approach than using Spack.

%
%
\bibliography{ref}

\end{document}